\documentclass[twocolumn,longbibliography]{revtex4-1}

\usepackage{graphicx}

\newcommand{\ket}[1]{\left\vert{#1}\right\rangle}

\begin{document}
\title{Novel constructions for the fault-tolerant Toffoli gate}
\author{Cody Jones}
\email{ncodyjones@gmail.com}
\affiliation{Edward L. Ginzton Laboratory,
         Stanford University,
         Stanford, California 94305-4088, USA}

\begin{abstract}
We present two new constructions for the Toffoli gate which substantially reduce resource costs in fault-tolerant quantum computing. The first contribution is a Toffoli gate requiring Clifford operations plus only four $T = \exp(i\pi\sigma^z/8)$ gates, whereas conventional circuits require seven $T$~gates.  An extension of this result is that adding $n$ control inputs to a controlled gate requires $4n$ $T$~gates, whereas the best prior result was $8n$.  The second contribution is a quantum circuit for the Toffoli gate which can detect a single $\sigma^z$ error occurring with probability $p$ in any one of eight $T$~gates required to produce the Toffoli.  By post-selecting circuits that did not detect an error, the posterior error probability is suppressed to lowest order from $4p$ (or $7p$, without the first contribution) to $28p^2$ for this enhanced construction.  In fault-tolerant quantum computing, this construction can reduce the overhead for producing logical Toffoli gates by an order of magnitude.
\end{abstract}
\maketitle

\section{Introduction}
Fault-tolerant quantum computing is the effort to design quantum information processors which are resilient to sufficiently small (but nonzero) probability of failure in any individual component~\cite{Preskill1998,Nielsen2000}.  Enhanced reliability comes at the cost of redundancy, and recent study in this area has focused on minimizing the overhead, or additional resource costs, associated with converting a perfect quantum operation into a form compatible with error correction~\cite{Isailovic2008,Jones2012,Fowler2012}.  This work focuses on the Toffoli gate, which appears in both reversible-classical and quantum logic and which may be defined as $\texttt{Tof}\ket{x,y,z} = \ket{x,y,z\oplus xy}$, for $x,y,z$ being binary variables.  Unlike many quantum gates, the quantum Toffoli gate has a classical analogue, so it is favored as a building block for importing more complex classical operations, such as binary arithmetic, into quantum algorithms like Shor's factoring algorithm~\cite{Shor1999,VanMeter2005,Jones2012} and quantum simulation~\cite{Clark2009,Jones2012,Jones2012b}.  For these reasons, the Toffoli gate is critically important to quantum computing in general, and improvements in the design of the Toffoli gate make the realization of large-scale quantum computation more tractable.

Several researchers have studied circuit constructions for the Toffoli gate.  The most oft-cited implementation is probably the one on page~182 of Ref.~\cite{Nielsen2000}, which may have been derived from Ref.~\cite{Barenco1995}.  As can be seen in Ref.~\cite{Nielsen2000}, the Toffoli gate is decomposed into smaller quantum gates, each of which can be made fault-tolerant by conventional means~\cite{Preskill1998}.  The most nettlesome of these is the $T = \exp(i\pi\sigma^z/8)$ gate, which is much more expensive in both time and space resources to produce~\cite{Zhou2000,Bravyi2005,Raussendorf2007,Isailovic2008,Jones2012,Meier2012,Bravyi2012,Jones2012c}; notably, the Toffoli circuit in Ref.~\cite{Nielsen2000} uses seven $T$~gates.  In fact, Ref.~\cite{Barenco1995} contains a construction nearly identical to one derived here (we use four $T$~gates), except for an undesirable $(-1)$ phase on one output state (we show how to correct this with modest effort).  However, ``complete'' implementations of the Toffoli gate without a phase error have used seven $T$~gates in the literature to date.  Amy~\emph{et~al.} studied classical search methods for decomposing gates like Toffoli into fault-tolerant primitives~\cite{Amy2012}, and Selinger investigated circuit constructions with particular emphasis on $T$-gate count and depth, where the latter metric allows parallel $T$~gates on different qubits~\cite{Selinger2012}.  We use Selinger's work as our starting point, as we turn his almost-Toffoli gate into a proper Toffoli gate, using four $T$~gates and some quantum teleportation.  Finally, the importance of this topic has attracted the attention of other researchers, and Eastin has independently discovered equivalent results~\cite{Eastin2012}.

This paper presents two important results.  First, Section~\ref{Four_T_Toffoli} describes how to implement the Toffoli gate with only four $T$~gates and Clifford-group operations~\cite{Gottesman1999,Nielsen2000}.  Second, Section~\ref{Error_detect_Toffoli} introduces a Toffoli construction requiring eight $T$~gates that can detect an error in any single $T$~gate.  This new circuit is an important development for fault-tolerant quantum computing, because it relaxes the requirements on high-fidelity $T$~gates that are expensive to produce; however, the circuit is probabilistic, and we discuss its proper usage.  Section~\ref{Analysis} presents some analysis of the resource costs and error rates for these circuits.  The paper concludes with a brief discussion of the impact these results have on large-scale quantum computing.

\section{Toffoli using just four $T$ gates}
\label{Four_T_Toffoli}
In fault-tolerant quantum computing, the most difficult quantum gates to produce are non-Clifford gates.  The Hadamard gate $H = (1/\sqrt{2})(\sigma^x + \sigma^z)$, the phase gate $S = \exp(i\pi \sigma^z/4)$, and the \texttt{CNOT} gate are generators for the Clifford group, as any gate in this group can be produced by combinations thereof, up to a global phase that we ignore.  However, at least one gate outside the Clifford group is required for universal quantum computing.  The $T$~gate is often selected because it is the easiest to produce; however, as we explain below, ``easy'' is relative, and this gate is still quite expensive in computing resources.

In most quantum codes, including the surface code~\cite{Fowler2009}, non-Clifford gates are produced using an ancilla state that is injected into the circuit~\cite{Gottesman1999,Nielsen2000}.  As this ancilla is produced in a faulty manner, it must be purified through magic state distillation~\cite{Bravyi2005,Meier2012}.  The handful of rounds of state distillation required to reach the $\sim 10^{-12}$ error rates required for quantum algorithms are considerably expensive, such that a single $T$~gate requires $\sim 100\times$ the circuit volume (product of qubits and time steps) of a \texttt{CNOT} or $H$~gate~\cite{Jones2012}, making its production the dominant cost among fault-tolerant gate primitives.  This poses an issue for quantum computing, as very many $T$~gates in the form of Toffoli gates are required for typical quantum algorithms like integer factoring or quantum simulation.  The first Toffoli gate construction we present uses four $T$~gates instead of seven, thereby reducing the overhead due to state distillation.

\begin{figure}
  \centering
  \includegraphics[width=8.4cm]{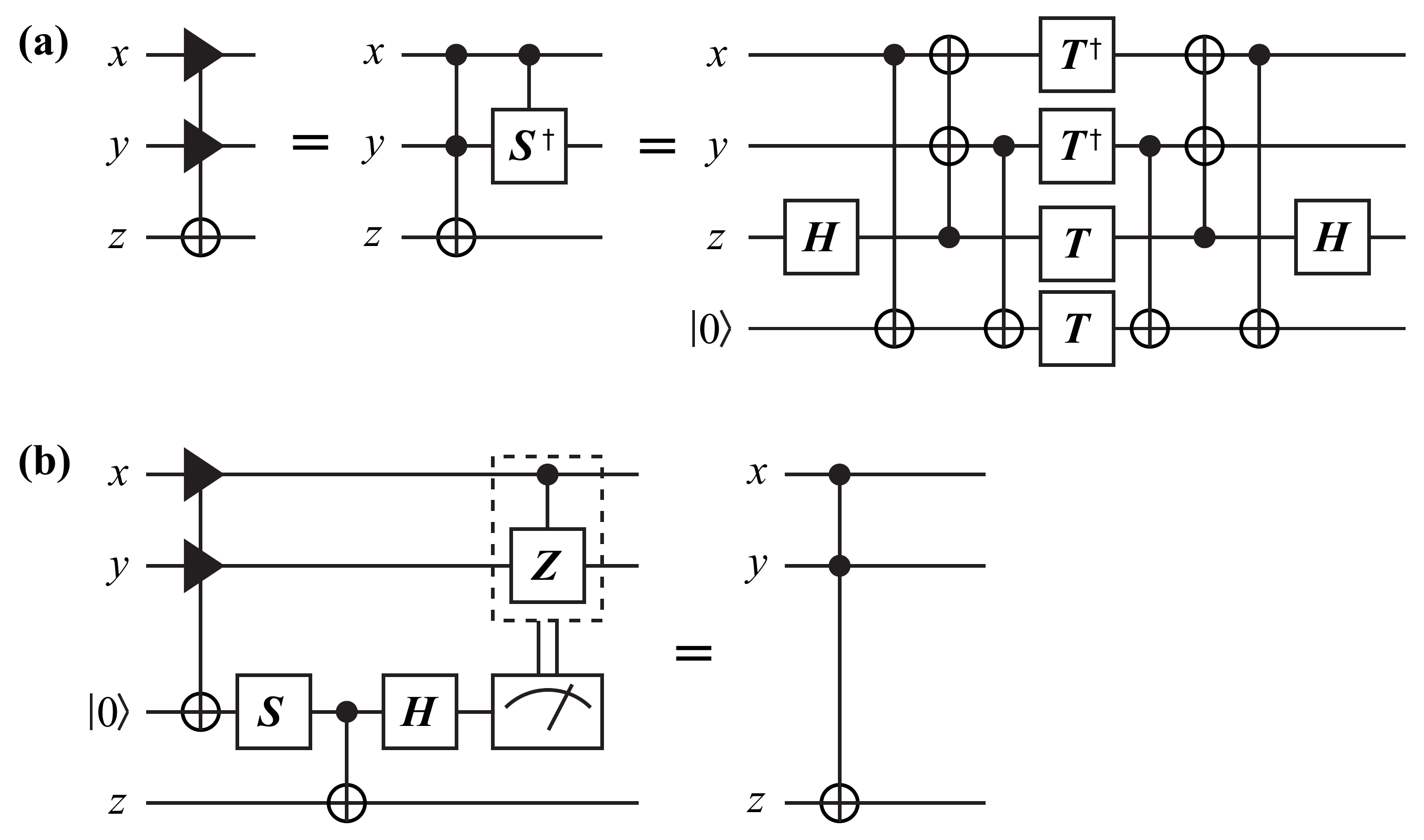}\\
  \caption{A circuit construction for Toffoli using four $T$~gates.  (a) The Toffoli$^{\star}$ circuit by Selinger~\cite{Selinger2012} that is almost a Toffoli, with the difference being the controlled-$S^{\dag}$ operation.  (b) Our circuit combines Toffoli$^{\star}$ with a phase correction and teleportation to produce an exact Toffoli gate.  The measurement is in the $\sigma^z$ basis, and the double vertical lines indicate that the controlled-$Z$ correction is conditioned on the measurement result being $\ket{1}$.}
  \label{four_T_circuit}
\end{figure}

Let us denote the Toffoli$^{\star}$ gate as the operation in Figure~\ref{four_T_circuit}a, which requires four $T$~gates and was introduced by Selinger~\cite{Selinger2012}.  Toffoli and Toffoli$^{\star}$ differ only by a controlled-$S^{\dag}$ gate between the control qubits $x$ and $y$.  Beginning with Toffoli$^{\star}$, we need only an ancilla qubit, a phase gate $S$, and teleportation to implement the exact Toffoli gate, as shown in Figure~\ref{four_T_circuit}b.  We first apply the Toffoli$^{\star}$ using the same controls the desired Toffoli but with an ancilla $\ket{0}$ as target.  The erroneous controlled-$S^{\dag}$ is corrected by a simple $S$~gate applied to the ancilla.  Afterwards, the \texttt{CNOT} and measurement teleport the doubly conditional \texttt{NOT} operation encoded in the ancilla to the target qubit of the desired Toffoli.  The measurement result determines whether a corrective gate of controlled-$Z$, which is in the Clifford group, is required to correct a $(-1)$ phase resulting from measurement back-action.  One can readily verify that only four $T$~gates are required in this procedure~\cite{Nielsen2000,Gottesman1999}.  Note that the inverse gate $T^{\dag}$ requires the same ancilla-based teleportation circuit as $T$, so these gates are equivalent in state-distillation cost and construction.

The construction in Figure~\ref{four_T_circuit}b can also be used to add control-qubit inputs to an existing controlled-$G$ gate, where $G$ is any unitary.  Replace the \texttt{CNOT} in Figure~\ref{four_T_circuit}b with controlled-$G$ (targeting however many qubits $G$ acts on), and the result is controlled-controlled-$G$.  By iterating this procedure, one can add $n$ controls to controlled-$G$ using $4n$ $T$~gates.  The best prior result required $8n$ $T$~gates~\cite{Selinger2012}.

\section{Error-detecting Toffoli circuit}
\label{Error_detect_Toffoli}
Whereas the previous section reduced the number of $T$~gates needed to make a Toffoli, this section addresses the resource-cost problem differently by making each $T$~gate less expensive.  The cost of a $T$~gate scales inversely with the probability $p$ of it having an undetected error, with a relationship where circuit volume (qubits $\times$ gates) is $O\left(\mathrm{poly}(\log(1/p))\right)$.  We introduce a new Toffoli gate that can detect an error in any one of eight $T$~gates.  As a result, the effective error probability of the Toffoli gate is $28p^2$ instead of $4p$ (we only consider lowest non-vanishing order throughout this paper since $p \ll 1$).  Even though twice as many $T$~gates are needed, they can tolerate larger error rates, so they are substantially less expensive to produce than would otherwise be necessary.

\begin{figure}
  \centering
  \includegraphics[width=6cm]{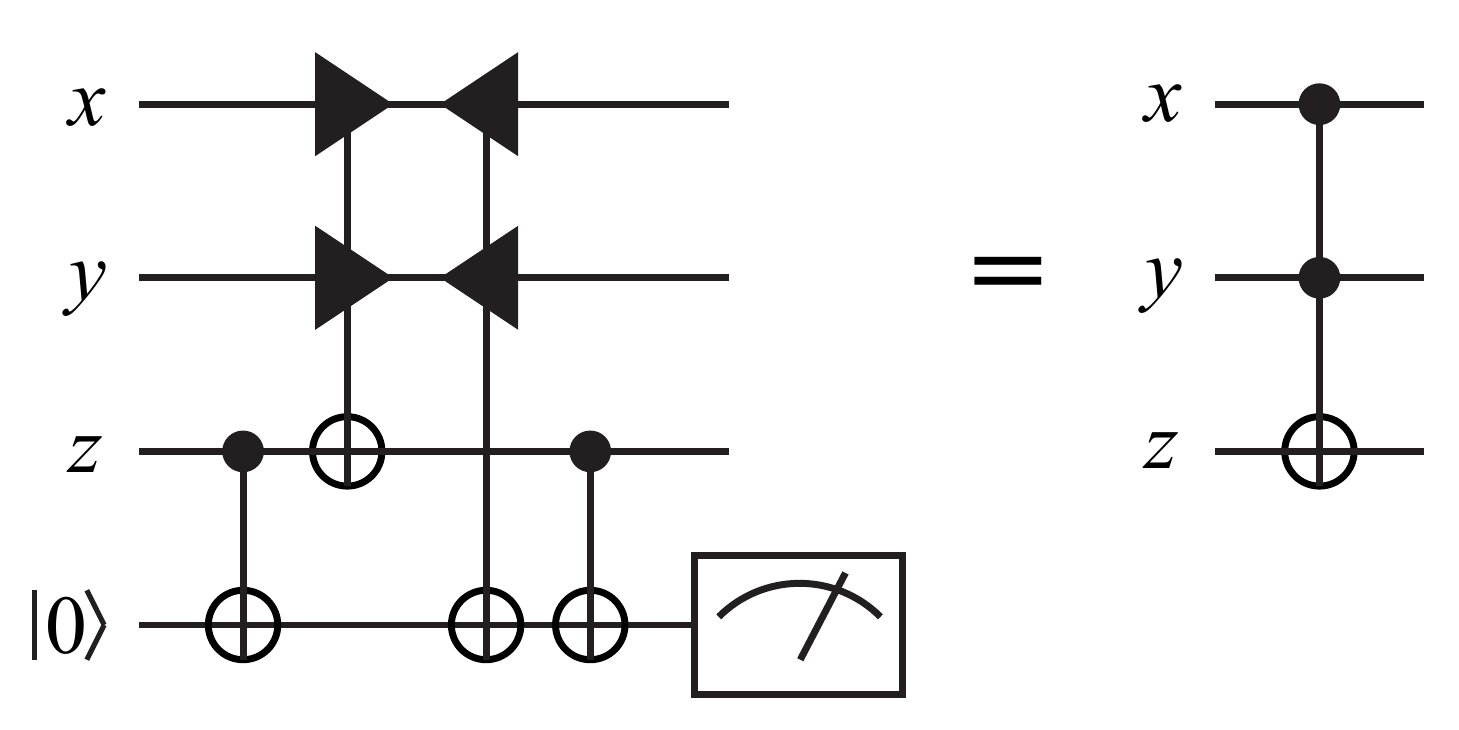}\\
  \caption{An error-detecting Toffoli gate.  The measurement is in the $\sigma^z$ basis, and obtaining result $\ket{1}$ indicates an error was detected, so the qubits should be discarded.}
  \label{error_detect_circuit_simple}
\end{figure}

\begin{figure}
  \centering
  \includegraphics[width=8.8cm]{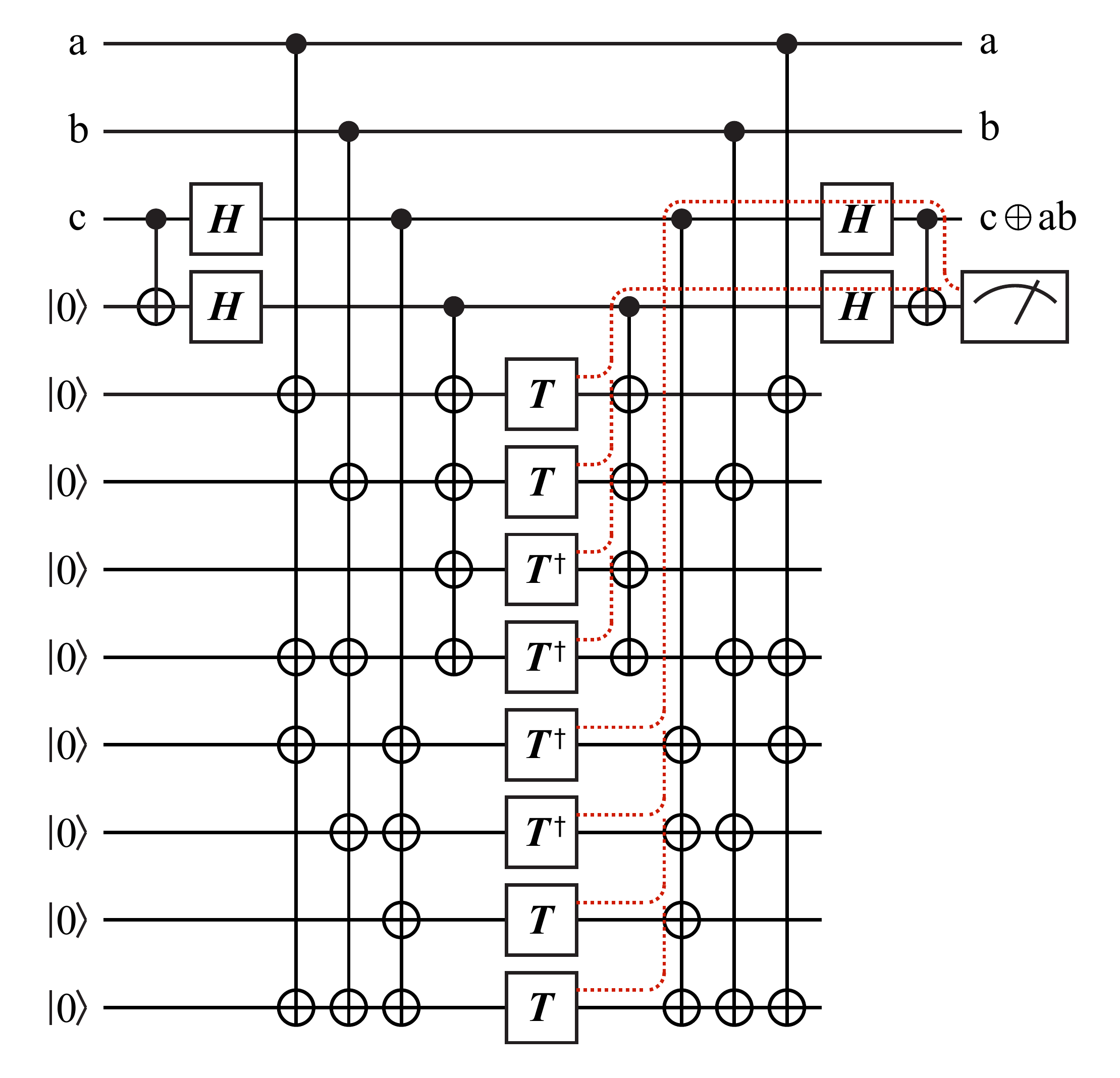}\\
  \caption{An error-detecting Toffoli gate.  The red dashed lines indicate how any single $\sigma^z$ error will propagate to the readout qubit.  The measurement is in the $\sigma^z$ basis, and obtaining result $\ket{1}$ indicates an error was detected, so the qubits should be discarded.  As long as the ancilla qubits are initialized perfectly to $\ket{0}$ and the \texttt{CNOT} and \texttt{H}~gates have no errors, then only $\sigma^z$ errors in the \texttt{T}~gates matter, as $\sigma^x$ errors cannot propagate to data qubits.  If the probability of a $\sigma^z$ error in each \texttt{T}~gate is i.i.d. $\texttt{Bernoulli}(p)$, then the success probability is $1-8p$ and the \emph{a posteriori} error probability is $28p^2$, to lowest order in $p$.}
  \label{error_detect_circuit}
\end{figure}

The error-detecting Toffoli circuit is rather simple to derive.  It consists of two Toffoli$^{\star}$ gates acting on a target qubit which is in a bit-flip code~\cite{Nielsen2000}, as shown in Figure~\ref{error_detect_circuit_simple}.  The gate with reversed triangles is the inverse operation (Toffoli$^{\star}$)$^{\dag}$.  Importantly, the controlled-$S$ and controlled-$S^{\dag}$ gates acting on the same qubits $x$ and $y$ are inverse operations, so they cancel.  A logically equivalent decomposition into $T$~gates is shown in Figure~\ref{error_detect_circuit}; this circuit is convenient for analyzing how errors propagate.  We assume that $\ket{0}$ preparation, $H$, \texttt{CNOT}, and measurement operations are perfect, because fault-tolerant error correction for these processes is economical compared to $T$~gates.  A single $\sigma^z$ error in any of the $T$~gates will necessarily propagate to the syndrome measurement for this bit-flip code, as indicated by the red dashed lines.  Upon such an event, all of the qubits are discarded.  Note that $\sigma^x$ errors, if present, do not propagate anywhere since they commute with the \texttt{CNOT} gates; they have no effect on the Toffoli gate.

\begin{figure}
  \centering
  \includegraphics[width=8.6cm]{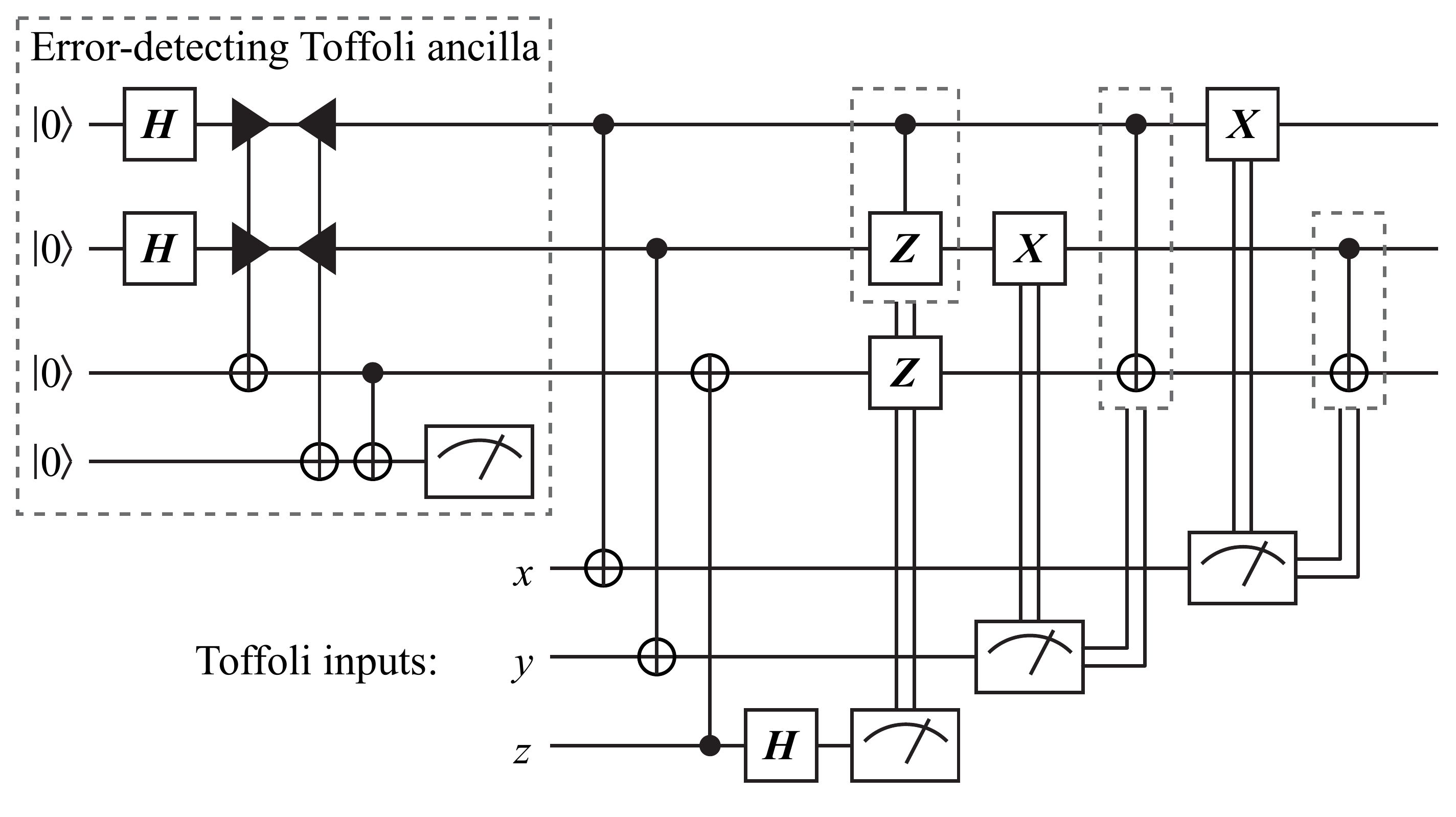}\\
  \caption{Proper use of an error-detecting Toffoli ancilla.  All measurements are in the $\sigma^z$ basis.  The ancilla-production circuit in the grey box in upper left is probabilistic.  A measurement result of $\ket{1}$ indicates the circuit failed, in which case the qubits are discarded.  Because the Toffoli ancilla is not coupled to any other part of the quantum computation, its production can be repeated until success.  The subsequent \texttt{CNOT} gates and measurements teleport data qubits through the Toffoli gate encoded in the ancilla (\emph{cf.} p.~488 of Ref.~\cite{Nielsen2000}).  Clifford-group gates are applied conditional on measurements showing outcome $\ket{1}$, as indicated by the parallel double lines.}
  \label{probabilistic_circuit}
\end{figure}

The circuit in Figure~\ref{error_detect_circuit} must be discarded upon a detected error event, which happens with probability $8p$.  If this circuit were connected by entanglement to other qubits in a quantum algorithm, all qubits must be discarded, and the algorithm fails.  To avoid this scenario, one can produce a Toffoli ancilla~\cite{Nielsen2000}.  If the circuit fails because of a detected error, then the qubits are discarded, but no far-reaching damage occurs since this faulty circuit is not entangled to any data qubits.  Conditioned on the circuit succeeding, the ancilla is teleported into data qubits to enact a Toffoli gate, using only Clifford gates and measurement, as shown in Figure~\ref{probabilistic_circuit}.  Using a representative value for $T$-gate error as $p = 10^{-8}$ (we consider such a scenario in Section~\ref{Analysis}), the failure probability for preparing the Toffoli ancilla is a modest $8 \times 10^{-8}$, which negligibly increases the number of times such preparation circuits must be repeated.

\section{Resource Analysis}
\label{Analysis}
Comparing resource costs between the naive Toffoli gate using seven $T$~gates and our construction using four is straightforward.  The latter requires about the half the resources of the former, under our assumption that $T$~gates are the dominant cost.  There is also a modest improvement in the Toffoli error rate ($7p$ becomes $4p$).  However, in fault-tolerant quantum computing, this result is likely overshadowed by the error-detecting construction.

Doubling the number of $T$~gates from four to eight to achieve $O(p^2)$ Toffoli error rate is usually the correct decision.  The reason is that this approach is more economical than increasing the accuracy of the $T$~gates through further magic-state distillation (or other fault-tolerant procedures).  Bravyi and Haah present a conjecture in the context of magic-state distillation, stating that to produce one magic state with error $O(p^2)$ requires at least two input states with error $p$; hence, the resources needed to increase $T$~gate accuracy to $O(p^2)$ at least doubles, and in all practical cases known to this author, the overhead factor is larger than two (an example case is considered below).  Moreover, there is no known protocol which saturates this bound.  Multilevel distillation comes arbitrarily close as $p \rightarrow 0$, but this limiting case is not always relevant for finite $p$, and multilevel protocols require large and complex circuits~\cite{Jones2012c}.

Under conditions relevant to quantum computing, the error-detecting Toffoli in Figure~\ref{error_detect_circuit} can reach the low error rates required for quantum algorithms with one less round of state distillation, leading to as much as an order-of-magnitude reduction in the resources required to produce a fault-tolerant Toffoli gate.  For example, suppose that we wish to produce a Toffoli gate with error probability below $10^{-12}$.  We presume the ``raw'' $T$~gate ancilla has a $\sigma^z$-error probability of $10^{-2}$.  Using the results in Ref.~\cite{Meier2012}, the simple Toffoli gate would require four $T$~gates distilled to $p = 10^{-15}$ using a hybrid scheme of one round of Bravyi-Kitaev (BK) distillation and two rounds of Meier-Eastin-Knill (MEK) distillation, at an average total cost of $1744.8$ raw states.  Conversely, the error-detecting Toffoli would require just one round each of BK and MEK distillation circuits for each of the eight $T$~gates distilled to $p = 10^{-8}$, for a total average cost of $697.6$ raw states.  The resource savings factor is $2.5\times$, just in terms of number of undistilled states needed for distillation.  In practice, the resource savings is amplified by another factor of 2$\times$ because one less round of distillation is needed (fewer gates means smaller circuit volume).  Additionally, the state-distillation sub-circuits for the error-detecting Toffoli gate can use weaker error correction (\emph{i.e.} lower code distance, by about a factor of two) than the same preparation circuits for the simple Toffoli, which translates to fewer qubits and gates at the hardware level~\cite{Fowler2012b}.  Relative to the Toffoli circuit using seven $T$~gates, there is an additional savings factor of $7/4$.  Therefore, the error-detecting circuit reduces total overhead for non-Clifford gates by up to an order of magnitude in this representative example.

It is also noteworthy that if ``raw'' $T$~gates can be produced with error rate $p=10^{-4}$, then the error-detecting Toffoli has a posterior error probability of approximately $3 \times 10^{-7}$.  This would enable modest quantum computations using about $10^6$ Toffolis, such as the multiplication of two 1000-bit numbers, without the need for resource-intensive magic-state distillation.

\section{Conclusions}
The Toffoli gate is an ubiquitous operation in quantum computing, as it plays a key role in many quantum algorithms.  However, quantum computers that realize these algorithms are still out of reach.  In the meantime, engineering a system capable of large-scale, fault-tolerant quantum computation demands that quantum computer architects minimize computing resource costs in terms of execution time and machine size.  The constructions in this paper substantially reduce the circuit volume for the fault-tolerant Toffoli gate when one considers how expensive each non-Clifford gate $T$ is to produce.  In the case of the error-detecting Toffoli gate, the resource savings is an order of magnitude in a representative example with $T$-gate error $p = 0.01$.  The improved fault-tolerant Toffoli gate brings large-scale quantum computing closer to realization.

\begin{acknowledgments}
This work was supported by the Univ. of Tokyo Special Coordination Funds for Promoting Science and Technology, NICT, and the Japan Society for the Promotion of Science (JSPS) through its ``Funding Program for World-Leading Innovative R\&D on Science and Technology (FIRST  Program).''
\end{acknowledgments}

\bibliography{References}

\end{document}